\documentclass[11pt,a4paper]{article}
\usepackage{graphics}
\usepackage{graphicx}
\usepackage{amssymb}
\usepackage{multicol}
\usepackage{cite}
\usepackage{amsmath}
\usepackage{amsbsy}
\usepackage{verbatim} 
\usepackage[usenames]{color}

\newcommand{\bEq}{\begin{equation}}
\newcommand{\eEq}{\end{equation}}
\newcommand{\bEQ}[1]{\begin{equation} \begin{array}{#1}}
\newcommand{\eEQ}{\end{array} \end{equation}}

\newcommand{\myVect}[1]{\pmb{#1}}

\begin{document}

\pagestyle{empty}

\null
\vskip 1.0cm
\begin{center}

{\Large  
{\bf On automatic determination of quasicrystal orientations \\
\vskip 0.4cm
by indexing of detected reflections}}\\

\vskip 2.0cm
Adam Morawiec
\vskip 0.5cm 
{Institute of Metallurgy and Materials Science, 
Polish Academy of Sciences, \\ Krak{\'o}w, Poland.
}
\vskip 0.4cm 
E-mail: nmmorawi@cyf-kr.edu.pl \\
Tel.: ++48--122952854, \ \ \  Fax: ++48--122952804 \\

 \end{center}

\vfill

\noindent
{\bf Synopsis} \\
Indexing of diffraction patterns
for application to 
automatic orientation mapping of quasicrystalline materials
is described.

\vskip 0.7cm

\noindent
{\bf Abstract}
\\
Automatic crystal orientation determination and 
orientation mapping are important tools 
for research on polycrystalline materials.
The most common methods of automatic orientation determination 
rely on detecting and indexing individual diffraction reflections.
These methods, however, have not been used 
for orientation mapping of quasicrystalline materials.
The paper describes necessary changes to 
existing software designed 
for orientation determination of periodic crystals 
so that it can be  applied to quasicrystals.
The changes are implemented in one of such programs.
The functioning of the modified program is illustrated by an 
example orientation map of an icosahedral polycrystal.

\vskip 0.7cm


\hfill \today

\newpage

\pagestyle{plain}

\section{Introduction}

Determination of orientations of crystallites, in particular for orientation mappings, 
is an important aspect of  
studies on polycrystals. 
However, data on orientation statistics of quasicrystalline  materials
are scarce. 
At present,  quasicrystals are not supported by 
the widely used fast commercial orientation mapping systems 
relying on detection of individual reflections and conventional indexing, 
i.e. on assigning indices to the reflections.
Therefore, electron backscatter diffraction
(EBSD) orientation maps of quasicrystalline aggregates have been
obtained by computer-aided manual indexing \cite{Tanaka_2016},
by matching experimental patterns to simulated patterns 
\cite{Singh_2019}
or to patterns obtained from a master reference pattern 
\cite{Winkelmann_2020},
and by automatic indexing using lattices of 
periodic approximants of quasicrystals
\cite{Cios_2020}; see also \cite{Baker_2017,Becker_2018,Leskovar_2018,Labib_2019,Labib_2020}.

The question is how difficult is the  
conventional indexing of quasicrystal diffraction patterns?
The general idea is simple and well known: 
one needs to replace the lattice basis by 
a frame or overcomplete set of 'basis' vectors \cite{Elser_1985}. 
However, in practice, 
there are opinions that indexing quasicrystal 
diffraction patterns is complicated. 
The methods described in 
\cite{Tanaka_2016,Singh_2019,Winkelmann_2020,Cios_2020}
are ways around the problem of conventional indexing of such patterns.
This paper demonstrates that the investment 
in adapting existing indexing 
software to solve quasicrystal diffraction 
patterns is relatively small.
The necessary changes are described in detail 
and implemented in one of the indexing programs.
The arguments are illustrated by an 
EBSD orientation map of an icosahedral polycrystalline material.

\section{From periodic crystals to quasicrystals}

\subsection{Orientations of periodic crystals}

It is worth recalling basic facts about  
orientation determination by indexing 
of detected reflections for periodic crystals.
Let $\myVect{s}$ denote a scattering vector normal to a reflecting crystal plane.
In  crystal diffraction, the scattering vector 
points to a node of the crystal reciprocal lattice,
i.e. it has the form of the integer combination 
$\myVect{s} = h \mathbf{a}^{\ast} +  k \mathbf{b}^{\ast} + l \mathbf{c}^{\ast}$,
where $hkl$ are reflection indices, and 
$\mathbf{a}^{\ast}$, $\mathbf{b}^{\ast}$, $\mathbf{c}^{\ast}$
are basis vectors of the reciprocal lattice.
With the  vectors $\mathbf{a}^{\ast}$, $\mathbf{b}^{\ast}$, $\mathbf{c}^{\ast}$ 
renamed to $\myVect{a}^1$, $\myVect{a}^2$, $\myVect{a}^3$, 
and the indices $hkl$ renamed to $l_1 l_2 l_3$, 
using the summation convention, the above expression takes the form 
\bEq
\myVect{s} =  l_i \myVect{a}^i \ . 
\label{eq:shkl}
\eEq
Besides having the basis vectors $\myVect{a}^i$
and the basis $\myVect{a}_i$ of the direct lattice, 
it is convenient to equip the crystal with a rigidly attached 
right-handed Cartesian system
based on vectors $\myVect{e}_i = \myVect{e}^i$ in which coordinates 
of $\myVect{s}$ are $s_i$, i.e. 
$\myVect{s} =  s_i \myVect{e}^i$.
Since the vectors $\myVect{e}_i$, $\myVect{a}^i$ and $\myVect{a}_i$
are known a priori, so are their dot products.
In particular, by definition, one has
$\myVect{a}_i \cdot \myVect{a}^j = \delta_i^{\ j}$,
where $\delta$ is the Kornecker delta.
Knowing the indices $l_1 l_2 l_3$ (i.e. $hkl$) of the reflecting plane, one can get
the Cartesian coordinates
\bEq
s_i  
= \myVect{s}  \cdot \myVect{e}_i
= l_j \myVect{a}^j  \cdot \myVect{e}_i  =  l_j B^{j}_{\; \, i} \ , 
\label{eq:sBl}
\eEq
where 
$B^{j}_{\; \, i} =   \myVect{a}^j \cdot \myVect{e}_i$
is the $i$-th Cartesian component of the $j$-th
basis vector of the reciprocal lattice.

Diffraction patterns are made up of 
traces of diffraction reflections.
Based on a position of a trace,
one computes the coordinates $s^L_i$ 
of the scattering vector $\myVect{s}$ 
in the right-handed laboratory Cartesian coordinate system
based on vectors $\myVect{e}_L^i$, i.e. one has 
$
\myVect{s} = s^L_i \myVect{e}_L^i 
$.
The coordinates $s_i$ and $s^L_i$ are related by 
\bEq
s_i 
= \myVect{s}  \cdot \myVect{e}_i 
= s^L_j \myVect{e}_L^j  \cdot \myVect{e}_i  = O_{i}^{\ j} s^L_j \ , 
\label{eq:s_O_sL}
\eEq
where $O_{i}^{\ j} = \myVect{e}_i  \cdot \myVect{e}_L^j$ are 
entries of the special orthogonal matrix $O$ representing 
the sought orientation of the crystal 
in the laboratory reference system.

In an experiment, a number of diffraction reflections are detected,
so one has the coordinates $s^L_i$ of vectors of a certain set, say $G$. 
On the other hand, 
there are numerous crystal planes which may lead to 
detectable reflections. Using 
indices of 
potential high-intensity reflections, one gets  
the coordinates $s_i$ of vectors of another set, say $H$.
The problem of orientation determination 
is to get the matrix $O$ relating 
(as many as possible) 
vectors from $G$ to some vectors from $H$.
The problem can be seen as matching 
the largest possible subset of $G$ to a subset of $H$. 
For descriptions of suitable algorithms 
see, e.g. \cite{Morawiec_2022} and references therein.

Details of how to calculate the coordinates 
of the vectors of the $G$ set 
depend on the diffraction technique, but generally 
the method is simply based on the definition of the scattering vector:
the vector is the difference of wave vectors of 
the reflected beam and the incident beam.

As for the vectors of $H$, 
one usually starts with 
indices of a single 
representative of each detectable family of symmetrically 
equivalent reflections, and then the coordinates of 
the scattering vectors corresponding to other reflections 
of the family are determined by using 
all symmetry operations for 
the crystal point symmetry: 
From the indices $l_{j}$ of the representative, 
one gets the coordinates $s_i = l_{j} B^{j}_{\ i}$ of the corresponding scattering vector, and 
with the orthogonal matrix $R$ representing a point symmetry
in the basis $\myVect{e}_i$, 
the coordinates of the equivalent vector are $R_{i}^{\ j} s_j$.\footnote{Care must be taken 
of cases when the vectors overlap with symmetry elements
and the number of distinct vectors is smaller than the number of symmetry operations.}

The integrity of the orientation determination procedure is 
confirmed by 
explicitly assigning indices to individual reflections.
To this end, the list of vectors in $H$ can be accompanied by a 
table with indices of vectors on the list,  
but a more convenient approach is to calculate the indices directly from the $s_i$ 
coordinates without creating any additional tables.
The indices are 
$l_i = l_j \, \myVect{a}^j \cdot \myVect{a}_i = 
\myVect{s}  \cdot \myVect{a}_i =  s_j  \myVect{e}^j \cdot \myVect{a}_i  = 
A_i^{\; \; j} s_j$,
where 
$A_i^{\; \; j} =   \myVect{a}_i \cdot \myVect{e}^j$
is the $j$-th Cartesian component of the $i$-th
basis vector of the direct lattice. 
If the coordinates are inaccurate,  as those computed  based on a symmetry operation 
or obtained from experimental $s^L_j$ via eq.~(\ref{eq:s_O_sL}),
the indices are 
\bEq
l_i = \left\lfloor   \myVect{s} \cdot \myVect{a}_i \right\rceil 
=  \left\lfloor A_i^{\; \; j} \,  s_j   \right\rceil  \ , 
\label{eq:indexing}
\eEq
where 
$\left\lfloor x \right\rceil$  
denotes the integer nearest to real $x$. 
If the magnitude of $\myVect{s}$ is not known, as in the case of 
scattering vectors corresponding to EBSD bands, 
one needs to test all admissible magnitudes.

\subsection{Orientations of quasicrystals}

The question is how the case of a quasicrystal differs from that of a periodic crystal.
As was already noted, 
the lattice basis $\myVect{a}^i$ ($i=1,2,3$) must be replaced by a frame, 
i.e. 
an overcomplete set of vectors $\myVect{a}^{\mu}$, where $\mu=1,2,\ldots , n \geq 3$ 
\ \cite{Elser_1985}. 
Every 
scattering vector can be expressed as a linear integer combination of the vectors $a^{\mu}$.
The expressions (\ref{eq:shkl}) and (\ref{eq:sBl})
need to be replaced by
$$
\myVect{s} = l_{\mu} \myVect{a}^{\mu}
\ \ \ \mbox{and} \ \ \ 
s_i 
 = \myVect{s}  \cdot \myVect{e}_i 
= l_{\mu} \myVect{a}^{\mu}  \cdot \myVect{e}_i  =  l_{\mu} B^{\mu}_{\; \; i} \ , 
$$
where $l_{\mu}$ are reflection indices
and 
$B^{\mu}_{\; \; i} = \myVect{a}^{\mu} \cdot \myVect{e}_i$.

The vectors $\myVect{a}^{\mu}$ correspond to the basis $\myVect{a}^i$ 
of reciprocal lattice, whereas,  
as a rule, the input of orientation determination systems dealing with periodic data
contains the basis of the direct lattice. 
To stay within this convention, 
one needs to input the frame $\myVect{a}_{\mu}$ dual to $\myVect{a}^{\mu}$. 
The set of vectors $\myVect{a}_{\mu}$ can be viewed as one 
whose subsets span quasicrystal tilings in physical space, i.e., 
vectors pointing to vertices of tiles can be expressed as linear integer combinations of the vectors $\myVect{a}_{\mu}$.

The Cartesian coordinates 
$B^{\mu}_{\; \; i}$ 
of the vectors $\myVect{a}^{\mu}$ are obtained from the input coordinates  
$A_{\mu}^{\; \; i} = \myVect{a}_{\mu} \cdot \myVect{e}^i$
of the vectors $\myVect{a}_{\mu}$ by using the generalized (Moore-Penrose) inverse \cite{Ben_Israel_2003}
of the transposed  matrix $A$, i.e.
$B = \left( A^T \right)^+ $. 
Clearly, if $n=3$, the vectors $\myVect{a}_{\mu}$ 
are linearly independent, 
the matrix $A$ is invertible, and $B$ is the regular inverse of $A^T$.

One also needs to recall that quasicrystals have the inflation/deflation property.
Unlike (Niggli reduced) bases of a periodic crystal lattices,
frames characterizing  quasicrystals and their diffraction patterns are not 
unique; they can be inflatated or deflated \cite{Elser_1985}. 
However, this is not an issue here because a specific frame $\myVect{a}_{\mu}$ is selected, and  
one only needs to ensure that the indices $l_{\mu}$ are correct for the dual frame $\myVect{a}^{\mu}$.

In the case of icosahedral quasicrystals,
it is convenient to use the frame of Bancel et al. \cite{Bancel_1985}
with $n=6$ and the vectors
\bEq
\begin{array}{ll}
\myVect{a}^{1} = \left( \myVect{e}^1 + \tau \myVect{e}^2 \right)/a \    &  \ \ \ \  
\myVect{a}^{2} = \left( \myVect{e}^1 - \tau \myVect{e}^2 \right)/a \   \\
\myVect{a}^{3} = \left( \myVect{e}^2 + \tau \myVect{e}^3 \right)/a \    &  \ \ \ \  
\myVect{a}^{4} = \left( \myVect{e}^2 - \tau \myVect{e}^3 \right)/a \   \\
\myVect{a}^{5} = \left( \myVect{e}^3 + \tau \myVect{e}^1 \right)/a \    & \ \ \ \  
\myVect{a}^{6} = \left( \myVect{e}^3 - \tau \myVect{e}^1 \right)/a \   \\
\end{array}
\label{eq:qc_frameBancel}
\eEq
along five-fold symmetry axes;
the vectors $\myVect{e}^i$ are along two-fold axes, 
$\tau$ denotes the golden ratio and $a$ is a structural parameter. 
The direct space frame $\myVect{a}_{\mu}$ 
dual to Bancel's frame $\myVect{a}^{\mu}$
is given by 
$\myVect{a}_{\mu} = \myVect{a}^{\mu} /(2(\tau+2))$. 
For alternative frames, see \cite{Elser_1985,Katz_1986} and \cite{Cahn_1986}.

Replacement of a lattice basis by a frame affects the generation 
of the theoretical scattering vectors of the set $H$. 
One complication is getting indices of symmetrically equivalent reflections.
Lists of equivalent indices for some quasicrystal symmetries are in 
\cite{Morawiec_2022}.
Moreover, the indexing 
based on eq.~(\ref{eq:indexing}) cannot be easily generalized to quasicrystals. 
A procedure 
described in section 13.5 of \cite{Morawiec_2022}
generalizes the conventional $l_i = \myVect{s}  \cdot \myVect{a}_i =  A_i^{\; \; j}   s_j$,
but it relies on 
distinction between rational and irrational numbers,
and it is inapplicable to inaccurate data.
An approach applicable to such data for icosahedral quasicrystals 
is in the Appendix.

All other aspects of quasicrystal orientation determination 
remain 
the same as for periodic crystals.
In particular, the method of calculating the measured scattering vectors 
of $G$ is the same for periodic crystals and quasicrystals.
As with periodic crystals, to get $H$,
it is standardly assumed that representatives of the families of reflections
that make up diffraction patterns are known a priori.
Finally, the method of matching the largest possible subset of $G$ 
to a subset of $H$ does not need to be changed.

\subsection{Modifications to indexing software}

The guidelines described in previous section were used to 
modify \textit{KiKoCh2} -- a program 
for orientation determination
via indexing of diffraction patterns, 
which was originally developed for dealing with 
periodic crystals. 
For a description of the original program, see \cite{Morawiec_2020}.

With $n$ denoting the number of frame-spanning vectors, 
the main change in the program is that the modified version 
allows for $n$ to be larger than $3$. 
The default value of $n$ (which is $3$) 
can be changed in the input. (See Table \ref{tab:1}.)
The other input data affected by this change are, first, 
the table $A$ with the coordinates 
$A_{\mu}^{\; \; i}$ 
of vectors $\myVect{a}_{\mu}$ of the direct space frame, and second, 
the lists of indices $l_{\mu}$ of representatives of families of reflectors
in the frame $\myVect{a}^{\mu}$.
With the dimension of the frame set to $n$,
the table $A$ 
consists of $n \times 3$ entries (instead of $3 \times 3$ entries for basis vectors), and 
the number of indices representing a family of reflectors is $n$
(instead of $3$).

\begin{table}
\begin{multicols}{2}

\noindent

\null

\vskip 0.6cm

{\small
\begin{verbatim}
_LatticeBasis
  1.0000000   0.0000000   0.0000000
  0.0000000   1.0000000   0.0000000
  0.0000000   0.0000000   1.0000000


_NumberOfFamiliesOfReflectingPlanes
       4
_FamiliesOfReflectingPlanes
       1   1   1
       0   0   2
       0   2   2
       1   1   3
 
_NumberOfSymmetryOperations
       24
_SymmetryOperations
  0.0000   0.0000   1.0000     0.00
  1.0000   0.0000   0.0000    90.00
 .......  .......  ........  ......
\end{verbatim}
}

\columnbreak

\noindent
{\small
\begin{verbatim}
_NumberOfBasisVectors
       6
_LatticeBasis
  1.0000000   1.6180340   0.0000000
  1.0000000  -1.6180340   0.0000000
  0.0000000   1.0000000   1.6180340
  0.0000000   1.0000000  -1.6180340
  1.6180340   0.0000000   1.0000000
 -1.6180340   0.0000000   1.0000000
 
_NumberOfFamiliesOfReflectingPlanes
       2
_FamiliesOfReflectingPlanes
       1   0   0   0   0   0
       1   1   0   0   0   0
 
_NumberOfSymmetryOperations
       60
_SymmetryOperations
  0.0000   0.0000   1.0000     0.00
  1.0000   0.0000   0.0000   180.00
 .......  .......  ........  ......
\end{verbatim}
}

\end{multicols}
\caption{
Parts of headers of input files for processing 
typical EBSD data for fcc metals (left column)
and 
for icosahedral quasicrystals (right column). 
The keywords used in the original version of \textit{KiKOCh2} have been 
left unchanged so that the modified program can process old data. 
The main difference is that the number of frame vectors 
is default $3$ in the left column, and it is
set at $6$ in the right column.
Consequently, 
the number of indices specifying families of reflectors is $3$
in the left column and $6$ in the right column.
Clearly, the two cases also differ by the number and type of point symmetry operations. 
}
\label{tab:1}
\end{table}

The only significant internal modification to the program 
concerns the calculation of the (reciprocal space) frame $\myVect{a}^{\mu}$
from the input (direct space) frame $\myVect{a}_{\mu}$. 
The subroutine for calculating the regular inverse of a matrix 
was replaced by a code for numerical computation of the Moore-Penrose inverse.

The modified software is universal in the sense that   
with $n=3$ it reduces to the original program for indexing 
data from periodic crystals. 
It is also applicable to periodic 
crystals with reflection indices specified in frames with $n$ larger than three
\cite{Morawiec_2016a}. 
In particular, it can be applied to data specified in 
hexagonal four-index setting or in quadray coordinates \cite{Her_1995,Comic_2016}.

The principles described above apply to diffraction patterns of various types.
In particular, they can be used to get orientations from patterns 
generated by EBSD. 
One only needs to take into account that with the usual EBSD band detection,
magnitudes of the scattering vectors are not available.
It just means that the vectors of both $G$ and $H$ are normalized to 1.

\section{Example} 

The performance of the program is illustrated on polycrystalline EBSD data. 
The data for suction-cast icosahedral TiZrNi are the same 
as those used in \cite{Winkelmann_2020,Cios_2020}. 
Diffraction patterns were collected 
using 
\textit{OIM Analysis}\textsuperscript{\texttt{TM}} software.
The software detects bands in the patterns by Hough transformation
and saves 
(Duda-Hart) line 
parameters corresponding to the 
bands 
\cite{Duda_Hart_1972}. 
These 
parameters were converted to normalized scattering vectors; 
for each line, its position was used
to get the coordinates $s^L_i$ of the unit vector
perpendicular to the plane 
containing the line and the point of origin of the pattern. 
Sets of the coordinates $s^L_i$ constitute the input of \textit{KiKoCh2}. 
The frame (\ref{eq:qc_frameBancel}) with $a=1$ was used. 
For the material under consideration, 
the strongest reflections belong to two families;
the scattering vectors of the first family are along five-fold axes, 
and the vectors of the second one are along two-fold axes.
Therefore, indices of representatives of the families 
were specified as $l_1 l_2 l_3 l_4 l_5 l_6  = 1  0  0  0  0  0 $ and $1 1 0 0 0 0$; cf. \cite{Bancel_1985}.
The input file also contained the 60
proper symmetry operations of the icosahedron.

\begin{figure}
	\begin{picture}(300,282)(0,0)
		\put(3,0){\resizebox{12.8 cm}{!}{\includegraphics{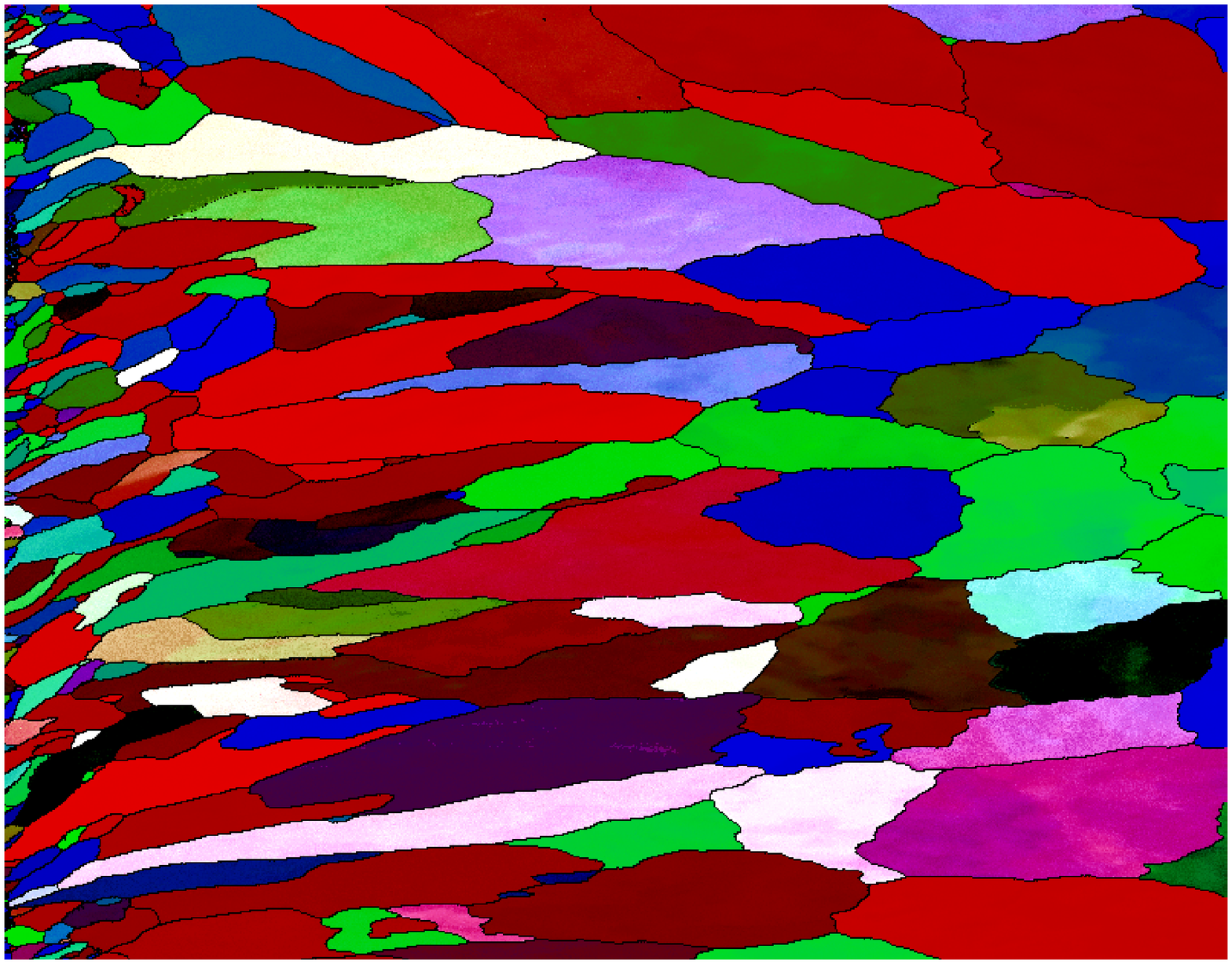}}}
		\put(379,0){\resizebox{1.5 cm}{!}{\includegraphics{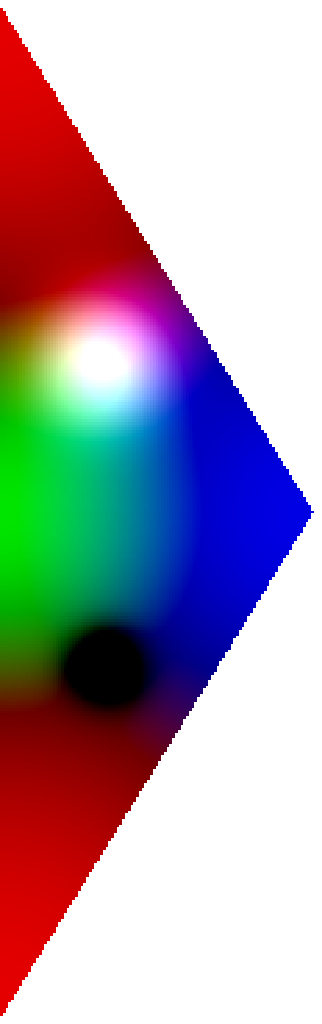}}}
		\linethickness{0.5mm}
		\put(377,282){\line(0,-1){96}}
		\linethickness{0.2mm}
		\put(374,282){\line(1,0){6}}
		\put(374,186){\line(1,0){6}}
		\put(380,231){50m$\mu$}
		\put(423,70){$\myVect{d}_3$}
		\put(378,140){$\myVect{a}_5$}
		\put(372,0){$-\myVect{a}_6$}
		
	\end{picture}
	\vskip 0.0cm
	\caption{
		Orientation map of icosahedral quasicrystal TiZrNi. 
		The coloring scheme is based on an arbitrarily selected direction.
		The triangle on the right is the domain of that direction.
		The vector $\myVect{d}_3$ is parallel to $\myVect{a}_1+\myVect{a}_5-\myVect{a}_6$,
		i.e. to one of the three-fold symmetry axes. 
		The map was not subject to any cleanup.
		Boundaries with misorientations exceeding $3^{\circ}$ 
		are marked in black.
	}
	\label{Fig_1}
\end{figure}

The file was processed by the modified \textit{KiKoCh2}. 
For almost all diffraction patterns,  
the number of detected bands was eight. 
Of the $799 \times 625 = 499375$ patterns, ten patterns were not solved;
in all these cases, the number of detected bands was 
smaller than three. 
With \textit{KiKoCh2}, the quality of an individual solution is quantitatively characterized 
by the number $N_u$ of indexed bands
and the fit $q$ of the detected and theoretical scattering 
vectors.\footnote{The fit is the arccosine of the average dot product 
of the detected and matching theoretical vectors.
The dot product of an individual pair of the vectors is 
$\delta^{ij} s_i O_{j}^{\; \; k} s^L_k = \cos \alpha$, where $\alpha$ 
is the angle between the vectors.
With $N_u$ pairs, 
one has $q = \arccos \left( \sum_{i=1}^{N_u} \cos \alpha_i/N_u \right)
\approx  \left(\sum_{i=1}^{N_u} \alpha_i^2 /N_u \right)^{1/2}$,
i.e. the fit is close to the root mean square of the angles $\alpha_i$.}
$N_u$ and $q$ depend on tolerances used for matching the vectors, but
with the default tolerances of \textit{KiKoCh2},
the average number of indexed bands 
was 7.66,
and the average fit 
was $0.78^{\circ}$.
The rate of indexing (serial computation on a 2.6GHz PC)
was more than $2.6 \times 10^4$ patterns per second. 
An additional small program for displaying the orientation map was written.
The resulting map is shown in Fig.~\ref{Fig_1}.
It is similar to those obtained in 
\cite{Winkelmann_2020} and 
\cite{Cios_2020}.

\section{Final remarks}

Automatic determination of crystallite orientations
by indexing detected diffraction reflections
is a fast and convenient tool for creating orientation maps
of polycrystalline materials. 
However, it has not been previously available for quasicrystals.
The paper describes modifications of software 
designed for periodic crystals that allow it to 
be used for quasicrystals.

The described modifications were implemented in the existing 
program \textit{KiKoCh2}.
The modified version of \textit{KiKoCh2} for Windows 
can be downloaded from \cite{myWebPageIndX}. 
The package also contains 
a short instruction for the program
and example data files. 
For illustration, 
\textit{KiKoCh2} was applied to indexing 
EBSD bands detected using a commercial EBSD system.
A clear orientation map of suction-cast TiZrNi icosahedral 
quasicrystal was constructed.

As has been with periodic crystals,  
the implementation of quasicrystal orientation determination 
in 
automatic orientation mapping systems will open other possibilities 
such as crystallographic texture determination, 
phase discrimination, 
determination of orientation relationships et cetera.

\vskip 1.0cm

\noindent
\textbf{{\Large Acknowledgments}}

\vskip 0.2cm

\noindent
I am very grateful to Grzegorz Cios 
for providing me with the file 
with 
positions of EBSD bands
and to 
Stuart~Wright 
for clarifying some details about this file.

\bibliographystyle{unsrt}
\bibliography{Icosah_map}

\newpage

\noindent
\textbf{\Large Appendix:  
Indices of computed scattering vector in 
 frame (\ref{eq:qc_frameBancel})}

\vskip 0.43cm

\noindent
Below is a procedure for determining reflection indices 
in the frame~(\ref{eq:qc_frameBancel}) of icosahedral quasicrystal
from approximate components of the scattering vector
given in the Cartesian system attached to the crystal.
Given the approximate coordinates $s_i=\myVect{s} \cdot \myVect{e}_i$, 
the task is to determine the indices $l_{\mu}$ such that 
$l_{\mu} \myVect{a}^{\mu} \approx \myVect{s} = s_i \myVect{e}^i$. 
The indices $l_{\mu}$ satisfy the relationship 
$$
A_{\mu}^{\ \, i} s_i=(\myVect{a}_{\mu} \cdot \myVect{e}^i) \, s_i = 
\myVect{a}_{\mu} \cdot \myVect{s}   
\approx  
\myVect{a}_{\mu} \cdot (l_{\nu} \myVect{a}^{\nu}) 
=  (\myVect{a}_{\mu} \cdot \myVect{a}^{\nu}) l_{\nu}
=  g_{\mu}^{\ \, \nu} l_{\nu} \ ,
$$
where $g_{\mu}^{\ \, \nu} =  \myVect{a}_{\mu} \cdot \myVect{a}^{\nu}$
are entries of a projection matrix.
With the frame (\ref{eq:qc_frameBancel})
and its dual $\myVect{a}_{\mu}$,
the explicit form of 
$g_{\mu}^{\ \, \nu} l_{\nu} \approx A_{\mu}^{\ \, i} s_i$
is
$$
\begin{array}{rcl}
\sqrt{5}  l_1 - l_2 + l_3 + l_4 + l_5 - l_6  & \approx & a \left(s_1/\tau + s_2 \right) \\
- l_1 +\sqrt{5} l_2 - l_3 - l_4 + l_5 - l_6  & \approx & a \left(s_1/\tau - s_2 \right) \\
 l_1 - l_2 +\sqrt{5}  l_3 - l_4 + l_5 + l_6  & \approx & a \left(s_2/\tau + s_3 \right) \\
 l_1 - l_2 - l_3 +\sqrt{5}  l_4 - l_5 - l_6  & \approx & a \left(s_2/\tau - s_3 \right) \\
 l_1 + l_2 + l_3 - l_4 +\sqrt{5}  l_5 - l_6  & \approx & a \left(s_3/\tau + s_1 \right) \\
-l_1 - l_2 + l_3 - l_4 - l_5 +\sqrt{5}  l_6  & \approx & a \left(s_3/\tau - s_1 \right) \ . \\
\end{array}
$$
This system of approximate equations can be 
solved with respect to integer $l_{\mu}$ in various ways.
One simple approach is to take 
the solution of the first, the third and the fifth equation 
with respect to 
$l_{2}$, $l_{4}$ and $l_{6}$
which is
$$
\begin{array}{rcl}
l_2 & \approx & 
l_5 + \tau (l_1 + l_3) - \xi_1 \\
l_4 & \approx & 
l_1 + \tau (l_3 + l_5) - \xi_2 \\
l_6 & \approx & 
l_3 + \tau (l_5 + l_1) - \xi_3 \\
\end{array}
$$
or 
\bEq
\ K_i \approx \tau L_i - \xi_i  \ , 
\label{eq:K_tL_xi}
\eEq
where $i=1,2,3$,
\bEq
\begin{array}{lcl}
K_1 = l_2 - l_5 &  \ \  &  L_1 = l_1 + l_3 \\
K_2 = l_4 - l_1 &  \ \  &  L_2 = l_3 + l_5 \\
K_3 = l_6 - l_3 &  \ \  &  L_3 = l_5 + l_1 \\
\end{array}
\label{eq:KL_l}
\eEq
and
\bEq
\begin{array}{l}
\xi_1 = a \left( s_1/\tau + \tau s_2 + s_3 \right)/2 \\
\xi_2 = a \left( s_2/\tau + \tau s_3 + s_1 \right)/2 \\
\xi_3 = a \left( s_3/\tau + \tau s_1 + s_2 \right)/2 \ . \\
\end{array}
\label{eq:xi_s}
\eEq
Knowing the coordinates $s_j$ ($j=1,2,3$), one determines $\xi_i$. 
The next step is to obtain the integers $K_i$ and $L_i$ satisfying the approximate relationship
(\ref{eq:K_tL_xi}).
This can be done by computing $\tau L_i - \xi_i$
for all $L_i$ with small absolute values $| L_i | \leq L_{limit}$,
and by choosing the pairs of $L_i$ and $K_i = \left\lfloor \tau L_i - \xi_i \right\rceil$ 
for which $\tau L_i - \xi_i$ is closest to an integer. 
Knowing $K_i$ and $L_i$, one obtains the indices $l_{\mu}$ 
by solving eqs.~(\ref{eq:KL_l}) or explicitly from 
\bEq
\begin{array}{l}
l_1 =  (L_1 - L_2 + L_3)/2 \\
l_3 =  (L_2 - L_3 + L_1)/2 \\
l_5 =  (L_3 - L_1 + L_2)/2  \\
\end{array}
\ \ \ \mbox{and} \ \ \ 
\begin{array}{l}
l_2 =  l_5 + K_1 \\
l_4 =  l_1 + K_2 \\
l_6 =  l_3 + K_3 \ . \\
\end{array}
\label{eq:l_LK}
\eEq
The procedure works only if the errors of $s_i$ and  
$L_{limit}$ are sufficiently small.
Therefore, in general, 
additional filters for rejecting unexpected sets of indices are needed.

It is worth illustrating the above scheme with a worked example.
Let $a=1$, and  
let the Cartesian components of the vector 
$\myVect{s}$
be 
$s_1 \approx -0.96$, $s_2  \approx 0.58$ and $s_3  \approx 1.63$. 
With these numbers, eqs.~(\ref{eq:xi_s}) lead to
$\xi_1  \approx 0.9876$, $\xi_2 \approx 1.018$ and  $\xi_3  \approx 0.01704$. 
The values of $\tau L_i - \xi_i$ for integer $L_i$ such that
$| L_i | \leq L_{limit} = 4$ are listed in Table \ref{tab:2}. 
\begin{table}
\begin{center} 
$$
\begin{array}{c|rrrrrrrrr}
L_1 &  -4 & -3 & -2 & -1 & 0 & 1 & 2 & 3 & 4  \\
\tau L_1 - \xi_1  & -7.46 & -5.84 & -4.22 &
   -2.61 & \mathbf{-0.99} & 0.63 & 2.25 &
   3.87 & 5.48  \\
\hline
L_2 &  -4 & -3 & -2 & -1 & 0 & 1 & 2 & 3 & 4  \\
\tau L_2 - \xi_2 & -7.49 & -5.87 & -4.25
   & -2.64 & \mathbf{-1.02} & 0.60 & 2.22 &
   3.84 & 5.45  \\
\hline
L_3 &  -4 & -3 & -2 & -1 & 0 & 1 & 2 & 3 & 4  \\
\tau L_3 - \xi_3 &  -6.49 & -4.87 & -3.25
   & -1.64 & \mathbf{-0.02} & 1.60 &
   3.22 & 4.84 & 6.46  \\
\end{array}
$$
\caption{
Values of $\tau L_i - \xi_i$ for
$\xi_i$ listed in the text and integer $L_i$ 
with absolute values not exceeding 4. 
For each $i$, the numbers $\tau L_i - \xi_i$  closest to integers are marked in bold. 
}
\end{center} 
\label{tab:2}
\end{table}
Based on the table, one has
$K_1= \left\lfloor -0.99 \right\rceil = -1$ and $L_1=0$, 
$K_2= \left\lfloor -1.02 \right\rceil = -1$ and $L_2=0$, 
$K_3=\left\lfloor -0.02 \right\rceil = 0$ and $L_3=0$. 
By using eqs.~(\ref{eq:l_LK}), one obtains the indices
corresponding to $\myVect{s}$; they are 
$ l_1 \,  l_2 \,  l_3 \,  l_4 \,  l_5 \,  l_6 = 
 0  \, \overline{1}  \, 0  \, \overline{1} \, 0 \, 0$.

\end{document}